# Power Off! Challenges in Planning and Executing Power Isolations on Shared-Use Electrified Railways

Alex Lu[1], Aleksandr Lukatskiy[1], Zhiqi Zhong[1], and John G. Allen, Ph.D.[2]

[1] Metro-North Commuter Railroad, P.O. Box 684, Ossining, New York 10562-0684.
Email: *lexcie (at) gmail.com* (corresponding author)

[2] Independent Transport Consultant, 5518 South Harper Ave., Chicago, Ill. 60637-1830

## Abstract

Electric railways are fast, clean, and safe, but complex to operate and maintain. *Electric traction infrastructure* includes signal power and feeder lines that remain live during isolations and complicate maintenance processes. Stakeholders involved in *power outage planning* include contractors, linemen, groundmen, power directors, dispatchers, conductor-flag, and support personnel. Weekly planning processes for *track time* requires many contingencies due to large number of moving parts and factors not known in advance, like personnel availability. Electrical and mechanical environments faced by crews working in adjacent areas may be entirely different and require a "bespoke" *circuit configuration* to *de-energize catenary*, which must be planned meticulously. Although recent automation improved real-time "*plate order*" communications between power directors and dispatchers, each outage still requires many manual *switching operations*. Net impact of this *isolation process* reduces available *construction work windows* nightly from a nominal 7 hours to 2 hrs 39 mins. We recommend joint design of electrical and civil infrastructure, cross-training between disciplines, limiting maximum number of concurrent outages, formal study of maintenance outage capacity, and further automation in power switching. We also believe costs arising from operation and maintenance of electrical infrastructure is a drawback that tips the balance towards diesel traction.

# Introduction

Much literature exists on the benefits of railway electrification. Electric railways are fast, clean, safe [1], modern, noiseless, can contribute to energy independence, and be quite sustainable (see e.g. [2]). Yet the cost and complexity of maintaining required electric traction infrastructure after initial construction seem to have barely entered investment policy discussions. This paper is a case study of processes and solutions on a mature, National Historic Engineering Landmark [3], high-voltage alternating current (AC) overhead electrified railway in overcoming complex challenges associated with keeping service operational whilst infrastructure maintenance and upgrade works are carried out. It also serves as a field guide for transport planners contemplating electrification as an option, providing some practical data on what should be expected in the operational phase of the project. At the same time, it provides a rare glimpse into the maintenance constraints that should be part of all service optimization models.

The primary motivation for this paper is to describe the processes involved in electric traction infrastructure maintenance, as a state-of-practice survey, and to provide a problem definition. We describe four key aspects of the power isolation process:

1. Work Planning
2. Circuit Configuration
3. Power Control
4. Switching Operations

We describe the electrical infrastructure only in general (and not in technical detail), as such information is well-documented elsewhere for the case study system [4] and in typical standard practice [5-8]. It does not attempt to solve the work schedule optimization problem or propose specific methods for automation implied by the findings; those are areas requiring further research.

## Infrastructure Basics

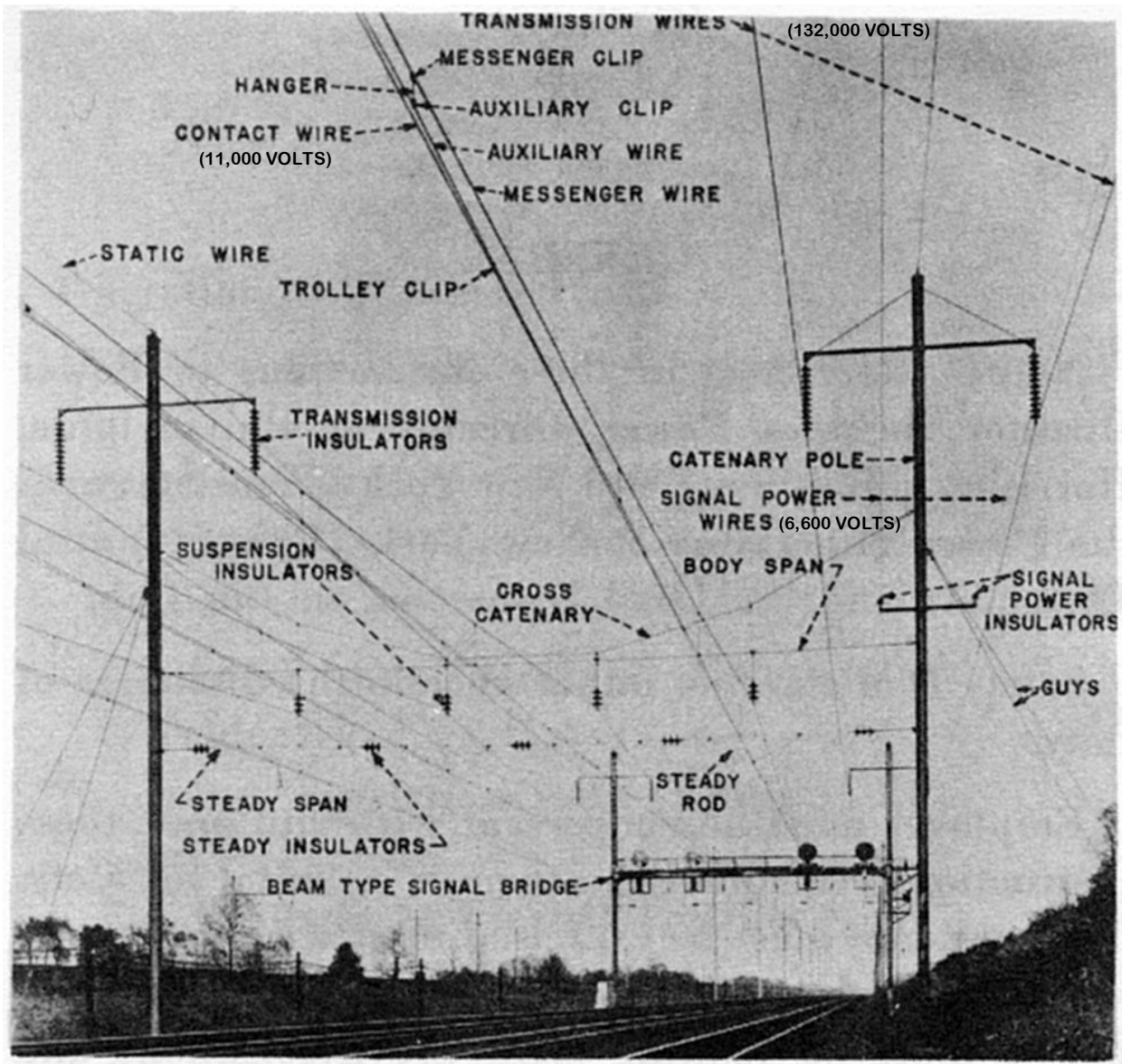

**Figure 1.** Typical catenary electrification infrastructure (Penn Central Railroad)

Figure 1 shows power line infrastructure broadly similar to those in the case study segment. This infrastructure set-up is typical of modern overhead electrified railways throughout the world, with minor variations [6]. Power infrastructure typically consists of at least four distinct portions:

1. **Trolley and Messenger Wires:** Typically thought of as "overhead line equipment" (OHLE), it provides the contact wire that train-borne pantographs come into contact with, to collect current.

2. **Power Feeder (and/or Return) Lines:** These carry bulk of the current for an OHLE segment to prevent signal interference and foreign earth problems (see e.g. [9]). Normally running at the same voltage (at opposite phase) with respect to ground as contact wires, they are connected to trolley wires at equalizing substations (also called autotransformers or "anchor bridges"). Typically carried at catenary height on the wayside, these lines must be grounded for maintenance work.

3. **Utility Supply Lines:** These are high-tension wires carried at much higher elevations above the catenary, or cross the right-of-way at strategic locations [5]. They provide three-phase power to supply substations and may additionally supply residential or industrial substations along the railway. Typically owned or operated by the local utility, they normally do not need to be grounded for railway work.

4. **Signal Feeder Wires:** These are medium voltage lines that supply signal substations along the railway to provide power to the signal system [10]. They may be underground or overhead and may carry power at a different voltage or AC frequency. These lines are particularly critical because signal power must be maintained even when overhead lines are grounded for renewal and substitute diesel service is in effect.

Power and signal feeders at catenary height abutting the right-of-way create complications for infrastructure maintenance and reconstruction. It is often necessary for some lines to remain live during work activities, complicating planning and execution, and limiting available work windows.

## Description of Current Processes

### Power Outage Planning

Railway operators are familiar with track outage planning (also called "possession planning") and the trials and tribulations that it entails (see e.g. [11,12]). On electrified railways, it is necessary to engage in power outage planning (i.e. "isolation planning") in conjunction with track outages, which significantly increases the complexity of planning, due to the larger number of groups involved, and the greater number of moving parts that can result.

### Planning Stakeholders

Many stakeholders are involved in setting up the possession and isolation plan for each week:

1. **Outside Contractors** needing outages to perform reconstruction or upgrade work

2. **Railroad Maintenance Personnel** (both within and outside the Power dept.) seeking outages to perform routine inspection and maintenance work

3. **Class A Linemen** required to protect Contractor work for electrical safety purposes, and to provide box grounds defining the limits of the isolation area

4. **Power Directors (Load Dispatchers, System Operators)** required to deliver on the planned power outage

5. **Substation Groundmen** required to assist the Power Directors in providing necessary grounds

6. **Rail Dispatchers** required to implement the planned track outage

7. **Conductor Flagmen** required to supervise Contractor work for track safety purposes

8. **Railroad Support Personnel** to support the possession and isolation plan, e.g. the Structures dept. erects "bridge plates" where trains are required to platform on the middle tracks, or the Signal dept. to remove parts of the signal system from service

9. **Railroad Management Personnel** perform various supervisory and coordination functions:

   a. **Capital Construction Inspectors** are required to inspect Contractor work (including any necessary engineering support);

   b. **Capital Trainmasters** are responsible for coordination between the Transportation functions and the construction project;

   c. **Project Managers** assume the coordinating role between all groups involved

10. **Sponsor Representatives** required to audit and oversee the entire project

11. **Local Utilities** supply high voltage lines to the substations. These are usually independent, but periodically require electrical outages to support necessary system repairs and improvements

12. **Foreign Railroads** operate their own electric trains over the case study segment, and need to be notified of isolation plans so they can organize alternate motive power, routes, or service plans

Generally, #9 and #10 do not introduce complications because they can carry out their work in an asynchronous manner. Because of the nature of the work required, #11 almost never interferes with other work directly, but may cause work rescheduling due to single-source feed contingency issues, which may limit the types of electrical outages granted. #12, foreign railroads, are considered as "customers" because they are users of shared infrastructure, so their concerns to protect their service must be addressed even though they do not directly perform maintenance or construction related work.

However, #1 through #8 have real-time resource requirements and must engage in real-time coordination. There are several challenges here. Class A Linemen, Substation Groundmen, Power Directors, Conductor Flagmen, and Rail Dispatchers are five separate crafts requiring different training and are drawn from different labour pools. On a given weekend, the effective capacity to support outages (and therefore to perform work) is constrained by the lowest availability of the five groups. If the plan requires 6 Linemen, 4 Groundmen, 2 Power Directors, 3 Flagmen, and 2 Dispatchers, but one of the Linemen bangs in sick right before the weekend, part of the planned work must be cancelled despite all other resources being available and ready to work.

Linemen and Groundmen perform two functions: routine maintenance (#2), and protective services (#3, #4). This results in competition for resources between performing in-house work and supporting Contractors, because personnel are drawn from the same labour pool. With a shortage of qualified craftsmen, no allocation scheme exists that totally eliminates project delays.

Contractors typically draw their construction labour resources from the local electrical workers' union. The work agreements for journeymen are similar to freelancers—allowing for generous leaves-of-absence and subject to reallocation between jobs. Railroad catenary reconstruction projects thus compete in real-time for skilled labour resources, from a limited supply pool, with other construction projects such as utility transmission lines, but also emergency projects like hurricane recovery tasks elsewhere in the country. Several Contractors work railroad reconstruction projects simultaneously, and thus even railroad projects compete amongst themselves for this limited supply of journeymen labour.

Due to the number of groups and the changing availability of resources (sometimes at short notice), it is critical to set up priorities that not only match the expected availability but also define which work arrangements will be cancelled if the anticipated resources do not materialize.  It is common to talk of having plans "A, B, and C" for each job, which defines contingency arrangements in case specific resources are unavailable when needed.

### Planning Processes

Based on this operational reality, specific processes are needed to both commit each stakeholder to supporting a specific plan, and to formalize the contingency plans defining what work to perform or cancel.  The planning process, typically organized by the project manager, generally works like this:

1. Each **Monday**, all Contractors submit a rolling "four-week look ahead" of planned work specifics, including required protective services.  Plans for the upcoming week are concrete, whereas the remaining three weeks are contingent upon progress during the upcoming week. Construction linemen on this railway generally work a Wednesday night through Monday morning cycle, with most critical work focused on Friday and Saturday nights.

2. On **Tuesday**, all Power stakeholders meet to define priorities of what work to cancel should the required resources not materialize, and define alternative work plans.  Internal railroad departments also attend this meeting to communicate their needs for power isolation.

3. On **Wednesday**, the formal "request" for support resources (Track Outages, Flagmen, Linemen, Groundmen, and Power Directors) is sent for approval by each responsible railroad department.

   a. When approving requests for Power Directors, Linemen, and Groundmen, Supervisors also consider the regular workloads within their own areas, and commit to providing resources only when not needed for regular maintenance.  Supervisors draft a personnel utilization plan, based on craftsmen vacation requests, shift patterns, expected vacancies, and overtime availability. The Supervisor balances the supply and demand for manpower at this point.

    b.   When approving requests for Flagmen, drawn from the same Extra Board as Train Service Conductors, Capital Trainmasters must also consider the needs of revenue train service. Capital Trainmasters also review the draft outage plan to ensure that it can be supported from a Transportation Field perspective (e.g. whether sufficient storage space is available in yards, whether the Mechanical dept. can supply the required number of sets, if extra fuel cycles are required on any substitute diesel equipment, etc.)

4.   On **Thursday**, the Transportation department meets to consider requests for track outages. Dispatching Chiefs consider the priority of work requiring outages for all departments and all Contractors, and their impacts on train service, including those of foreign railroads. The conflicting needs of work windows and service impacts are balanced at this point. Dispatching Chiefs check with each responsible railroad department to see if the outage plan can be supported, e.g. Structures department erecting bridge plates from high-level station platforms across tracks being removed from service, and consider the availability of Dispatchers in appropriate Districts, including their availability for overtime. The approved possession plan informs the isolation plan and may result in certain work being cancelled, and contingency plans being put in effect.

5.   On **Friday** (and sometimes, late on Friday, after execution of the weekend plan has already started) is when several real-time factors will definitively be known:

    a.   Last-minute cancellations or workplan changes by Contractors owing to factors such as contract labour unavailability or equipment malfunctions.

    b.   The Power department may initiate cancellations due to personnel shortages for reasons such as anticipated overtime not materializing, or employees taking sick leave. (Labour agreements govern how much notice is required when a Lineman is unwilling to fulfill an overtime request.)

    c. The Power department may also cancel work due to upcoming severe weather conditions (such as high winds), which may cause significant damage to overhead electrical wires.

    d. The Transportation department or any Maintenance of Way area may impose cancellations due to the needs of Train Service. Typical reasons include unexpected incidents during the Friday evening rush hour, or major unanticipated infrastructure failures requiring emergency repairs, which may affect both the available track outages (due to routing traffic around the affected area) and skilled personnel (for unplanned repairs).

Although fraught with points requiring short-notice revisions to weekend isolation plans (and thus the entire work plan), this planning process actually works well if project managers are proactive, and stakeholders stick with it. The process is, by nature, very decentralized, requiring active coordination across organizational boundaries. However, multiple touch points exist within each group's workflow, so a different structure probably would not deliver much benefits.

The key to efficient outage planning is for all stakeholders to actively support the process and be fully engaged. Although stakeholders can attempt to control variables within their domain, there will necessarily be last-minute changes. Trust and good working relationships between the personalities in charge are important; it is owing to these relationships that the railroad is able to accomplish this work.

Railroads have historically dealt with unscheduled employee absences by providing an "extra board" of craft employees who can be assigned to any job at a moment's notice, who may receive guaranteed pay even if not called for work, for essentially protecting jobs when the scheduled resource becomes unavailable at short notice. This practice has not historically extended to maintenance and engineering departments where any unavailable employee meant a specific job is "blanked" or uncovered. For intricate work plans, the benefits of carrying this extra backup resource could be substantial, although in practice it is essentially equivalent to making more conservative assumptions about employee availability when planning work.

Some railways have a more formal process to deal with track and power outage planning, requiring sign-offs at a high-level each week, and for last-minute changes to the plans. The case study system has a policy of accommodating these changes whenever possible. A more formal system would require longer planning horizons, and result in delays to jobs due to reduced nimbleness and agility. Requiring higher resource availability will in effect result in more resources being held in reserve, decreasing utilization.

**Power Isolation Circuit Configurations**

Most rail professionals have some experience with routing trains around a track obstruction, making moves using appropriate combinations of signals, crossovers, tracks, and alternative routes. Although isolating a segment of overhead catenary is similar in concept, a much more detailed level of operational consideration for wayside infrastructure is required. Planned power outages must:

1.  Provide electrical paths to allow power to be delivered outside the de-energized area, as it relates to trolley wire on unaffected tracks, feeder power lines, and signal power.

2.  Keep contact wires over crossovers, drawbridges, and other strategic infrastructure energized whenever possible, to maximize routing flexibility and minimize the risk of stranding electric trains.

3.  Ensure that power supply does not become "unbalanced", i.e., overloading specific supply or equalizing substations while leaving available capacity underutilized.

Power infrastructure is generally designed to make #2 easy to accomplish, by strategic placement of motorized disconnects (MODs), air gaps, and sectionalizing insulators (SIs), e.g., Figure 2. However, Power Directors face the opposite challenge from that faced by Dispatchers. To allow a critical interlocking to remain live for traffic crossing over during an outage, it requires catenaries within the interlocking to be capable of being fed from both ends. This adds to circuit complexity in a number of switches, and cross-feeding.

Substations are not typically located exactly at interlocking locations, due to engineering reasons including space requirements for modern substations, maximum distance between substations due to wire resistance and inductive reactance, and other electrical considerations (see e.g. [8,14]).

Interlocking locations are determined based on train service needs and physical site characteristics, whereas substation locations are selected based on power demand requirements and power supply convenience [15]. This introduces circuit complexity within the wire network to accommodate the switching and isolating requirements relating to the substations themselves.

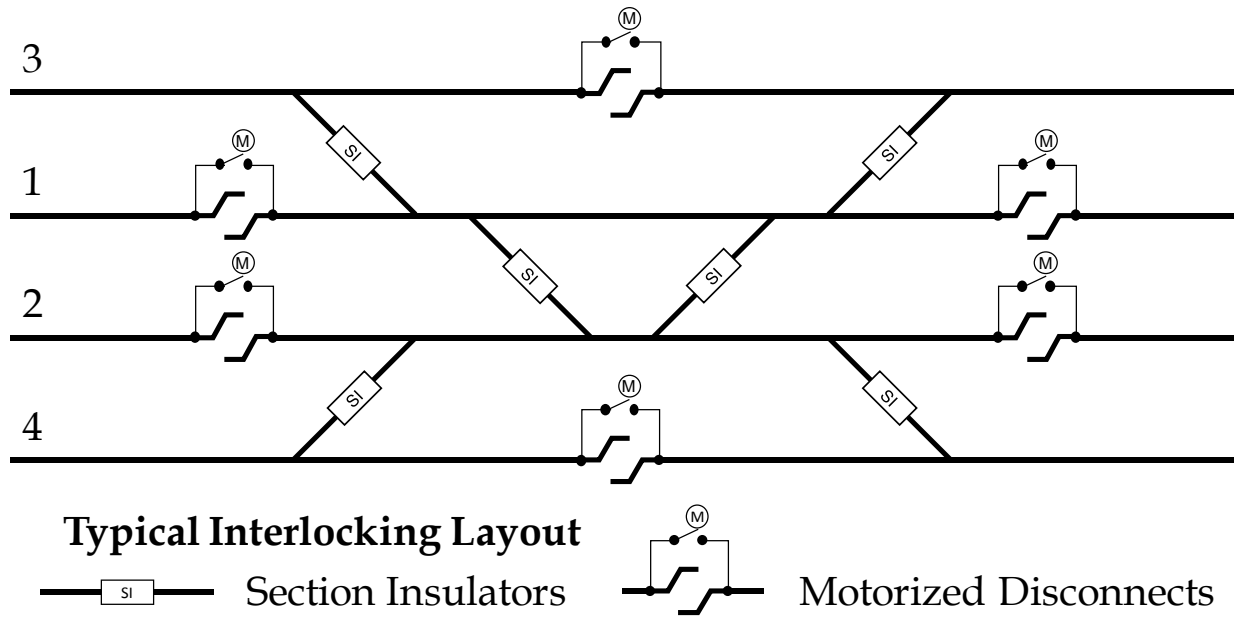

**Figure 2.** Typical power infrastructure layout at full universal interlockings

Catenary and feeder lines form a mini power grid that supplies power from substations to mobile loads (i.e., train locomotives drawing power). However, it also forms links in the wider utility grid due to interconnections via supply substations. Utility grids are typically geographic in nature, with natural boundaries such as rivers and glens dividing the power infrastructure into zones synchronized to the same phase timing. The railway breaches these geographic boundaries by electrified river bridges. Specific electrical infrastructure (phase breaks) are needed to deal with this. When it becomes necessary to feed power across a phase break during an outage condition, precautions must be taken to prevent different utility grids from becoming electrically connected.

This discussion sets aside the 3.2-km (two miles) maximum length of a "wire run" due to physical tensioning limitations in the "constant tension" catenary [16]. At the end of each run, the wire is terminated at a pull-off pole on the wayside via counterweights and has associated physical infrastructure (e.g. C-jumpers, cross-droppering, and overlapped catenary). Although continuous from an electrical standpoint, these are important locations from a mechanical standpoint and affect the feasible placement of electrical components. They also require regular maintenance and are special items to be worked on during renewal activities, and affect outage requirements (see e.g. [7], for details). Another constraint relating to configuring power outages appropriately is that the maximum work zone on the case study system is 30 catenaries (2,743.2 m (9,000 ft)) long. This is the maximum distance where it is safe to "box ground" the zone at both ends without induced voltages (sometimes called "static electricity") becoming a hazard.

These constraints result in a high degree of customization being required to configure circuits for wire work at any specific location. Although the crew may be working on the same track and the same signal block a few catenaries apart on two consecutive nights, the electrical and mechanical environments they face may be entirely different and may require a different circuit set-up to provide the necessary isolation. Each outage in effect requires a "bespoke" circuit configuration and must be planned meticulously by Power Directors, over-and-above any workload generated by the outage planning processes discussed previously.

**The Power Director–Train Dispatcher Interface**

Train Dispatchers operate controlled signals, power switches, and electric locks remotely using the Centralized Traffic Control (CTC) system; they may also operate dark (non-CTC) territory by issuing paper authorities to field personnel to order train or switch movements. Dispatchers have a variety of job aids or systems that provide for partial automation of these tasks, like Entrance-Exit (NX) controls, and interlocking logic that prevent governing signals from being cleared when switches are not properly lined, or when switches on opposite ends of crossovers are out of correspondence. These systems are classified as industrial control systems and provide a high degree of safety by providing system checks against possible human errors.

Similarly, Power Directors ("System Operators" in utility parlance) operate the power distribution system remotely using practically identical technology called Supervisory Control and Data Acquisition (SCADA) systems; they also order field breakers to be operated or lines grounded by issuing operating orders to Linemen in the field who fulfill these orders (e.g. [17,18]). The state of the art in automation of power system operating tasks is not quite as highly developed as those governing train movements. Power Directors typically make circuit changes by commanding operation of individual disconnects or tie breakers in a specific sequence, whereas a Dispatcher can "line a route" and have all affected switches and signals operated automatically. The SCADA system is therefore more manual, relies far more on Power Directors' paper checklists rather than system logic to ensure safe operations, but it is also more flexible in terms of how the circuit can be (validly) configured. Power Directors are personally responsible for the correct sequencing of each operation; no "interlocks" exist in the computer that prevent, e.g. an energized circuit from being connected to a grounded one, if a switch between the two lines was closed in despite one side being grounded. To prevent errors or oversights from leading to accidents, elaborate procedures and recordkeeping requirements were developed as administrative controls. This limits the throughput in terms of how quickly individual outages can be set up, and makes it all but impossible to delegate planning of switching sequences. Even if pre-printed lists of operations were made available, because Power Directors are personally responsible for each breaker operation, they must verify the system state prior to commanding each circuit change.

Despite utilizing independent control systems having different philosophies of operation, the actions of Power Directors and Dispatchers at the same physical locations must line up. If a segment of catenary is grounded for an outage, then the signals and switches must be set up such that no electric trains are permitted to enter the segment. (If the tracks are also out of service, no trains at all can be allowed to enter.) Because some equipment has multiple pantographs bussed together, allowing an electric train to travel across an air gap or SI between energized and de-energized sections momentarily allows two sections to be electrically connected, resulting in backfeeding through the high voltage bus onboard the train. This can cause equipment damage if the de-energized section was properly grounded; if not, this causes a dead section to become hot momentarily and is very dangerous. Electric trains must not enter dead catenary sections.

Interfaces between Power Directors and Dispatchers are governed by Plate Orders (e.g., Figure 3). These are pre-defined sets of limits delineated on a diagram showing track segments that cannot be used for electric operations, and identifies switches that must be blocked unless a diesel train wishes to use them [19]. Typically, the area having electric operations barred (running from switch to switch) is slightly larger than the actual segment of catenary being isolated (from SI to SI) as a safety backstop. Plate Orders must be pre-defined, published, and agreed upon by the responsible Power and Signal Engineers, to minimize risks of real-time confusion. They also limit the combinations of switches that can be utilized as outage limits, therefore permutations of outages that can be practically granted even though each control system may be more flexible. The case study territory has 200 published plate orders, including some very complex ones around storage yards. Identifying those closest to the specific working limits required by construction can become a limiting factor on how renewal or maintenance work must be sequenced, and hence on the maximum scope of work that is feasible to complete within one night's work window.

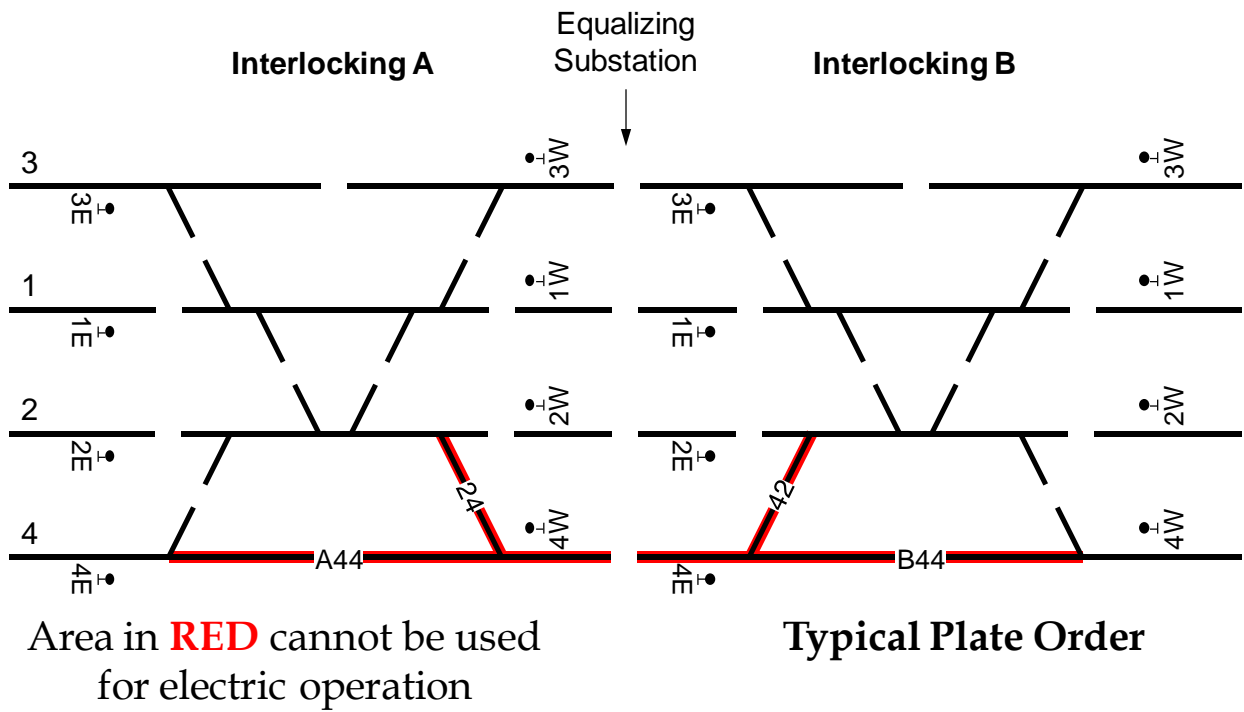

**Figure 3.** Typical plate order restricting electric train operations during power outages

Recently, the communications process between Power Directors and Dispatchers has been improved. Power Directors can now request any given plate order be put in effect by Dispatchers in the relevant district by selecting it on the Plate Order Protection System (POPS), which causes a dialogue box to pop up on the CTC screen requiring the Dispatcher to acknowledge that the plate order is now in effect. The CTC system then electronically locks out all relevant switches and signals and prevents them from being operated [20]. This innovation has reduced the time taken to implement plate orders and more importantly, improved safety by reducing the possibilities of human error.

Some electric railways have solved the Dispatcher-Power Director interface problem by combining the two functions. On an overhead DC electrified line in the Midwest, Load Dispatchers or Load Supervisors have been responsible for the operation of trains and electrical infrastructure alike [21] since the start of electric service in 1926.

Indeed, prior to modernization, autotransformer substations on the case study segment were co-located with signal switch towers and locally controlled by Tower Operators who were paid a supplementary hourly rate for assuming the additional responsibility of operating the power board, although on this system Load Dispatchers and Train Dispatchers always were separate crafts [22] who communicated with Tower Operators by phone. The 1986 conversions to 60Hz commercial utility electrical feed (from the original 25Hz system) and the 1996 conversion to CTC effectively severed the relationship between electrical and signal infrastructure (and operating personnel) in the field.

**Traction Power Switching Operations**

Typically, four steps are involved in providing electrical outage:

1. Receive the request (in real-time) from the field;

2. Command the change (via SCADA or paperwork);

3. Physically execute the change (e.g., operating an in-line knife switch, motoring an MOD, racking out a breaker, attaching grounds, etc.);

4. Tag the device with information about the circuit change and the person responsible, to prevent re-setting in error. This also includes SCADA software tags for remotely-enabled devices.

For a fully remote-controlled device accessible via the SCADA system, this sequence may take only 30 to 90 seconds to accomplish, where the tagging is done electronically on the SCADA system. However, where physical operations are needed in the field, the sequence will take much longer to execute. The Power Director typically makes a phone call to a mobile device carried by the Linemen (or a substation phone) to issue a paper order. Linemen are required to read it back and complete their own copies of paperwork in the field. Once the order is clearly understood, the Linemen can drive or walk to the nearby physical location to execute the change [23]. Tagging in this case is done physically with a tag attached to the device. Then the Linemen must communicate to the Power Director that the order has been fulfilled.

### *Number of Operations*

Although commanding a single circuit change may seem quick and simple enough, how many circuit changes does it take to implement a typical power outage? Normally these (conceptual) steps are needed to isolate a segment of electric catenary:

1. De-energize a large area by opening appropriate breakers at nearby equalizing substation(s).

2. Normally the limits of isolation are smaller and not exactly identical to the area supplied by one single substation, which requires operation of MODs and/or in-line switches to reduce outage limits. Open all disconnects (normally closed) that connect the isolated segment to adjacent areas. These switches cannot be opened under load, to reduce the risk of arcing.

3. If this arrangement de-energizes any adjacent sections that should remain live due to the local circuit layout, close appropriate switches to provide backfeed to affected segments.

4. If a substation is present in the segment and must remain dead, open all breakers feeding the supply taps, including breakers on the feeder side if applicable. Apply system grounds to the supply taps and rack out breakers if necessary.

5. Regular grounds are installed in the track clearance envelope directly between the contact wire and the running rails, to clearly indicate that the segment is out of service. If it is necessary to install aerial grounds (as to allow a diesel train to travel through the dead segment without interference), it will be necessary for Linemen to climb the poles. It may then be necessary to de-energize additional segments to allow the Linemen to work on the catenary structure, and then re-energize them after the work is complete.

6. If there is an anchor bridge in the segment that needs to be grounded, it will be necessary for safety reasons to momentarily de-energize a large area while work is carried out on the anchor bridge.

7. When the circuit is in its final outage configuration (including all the appropriate backfeeds), the entire area is re-energized by closing appropriate substation breakers. This action should leave the limits of isolation in a de-energized state and everything else energized.

8. Linemen in the field travel to the isolated segment, test for potential, and install box grounds to "box in" the work site. Box grounds protect workers in the outage by ensuring a zone of equipotential between the two grounds, in case of accidental re-energization from either end.

We will look at one specific real-life example in a suburban area where Contractors had requested a four-track power outage for bridle extension removal, a fairly typical work scope in catenary renewal. Power outage across all four tracks were required because catenary pull-offs are anchored on wayside poles and the wires cut through every track at some point within the outage limits. Although four-track outages sound extreme, they are actually routinely required and there may be 1-3 such instances every weekend. This specific example is from a project replacing historic fixed-termination equipment with auto-tensioning catenary, but similar switching operations are required for any catenary inspection, maintenance, or steady-state renewal work requiring power isolations.

One anchor bridge (awaiting final decommissioning), one malfunctioning substation, and one movable river bridge all fall within the limits of this four-track outage. Five movable bridges, 20 equalizing substations, 4 supply substations, and several anchor bridges are packed into this 89.6-km (56 mile) segment of electrified railway, so it is typical to find these "obstructions" while trying to set up an outage required for specific work scopes. (An atypically large number of equalizing substations exist due to a 1986 project to convert from 11kV 25Hz operation to 12.5kV 60Hz, resulting in increased inductive reactance in the catenary [24].) On this evening, another planned isolation is working nearby, which required two Power Directors to share control of some specific switching equipment (known locally as "double-header"), which increases coordination requirement and paperwork burden to ensure safe working. Once again, due to the large number of renewal projects currently working on this railway, this is a normal occurrence.

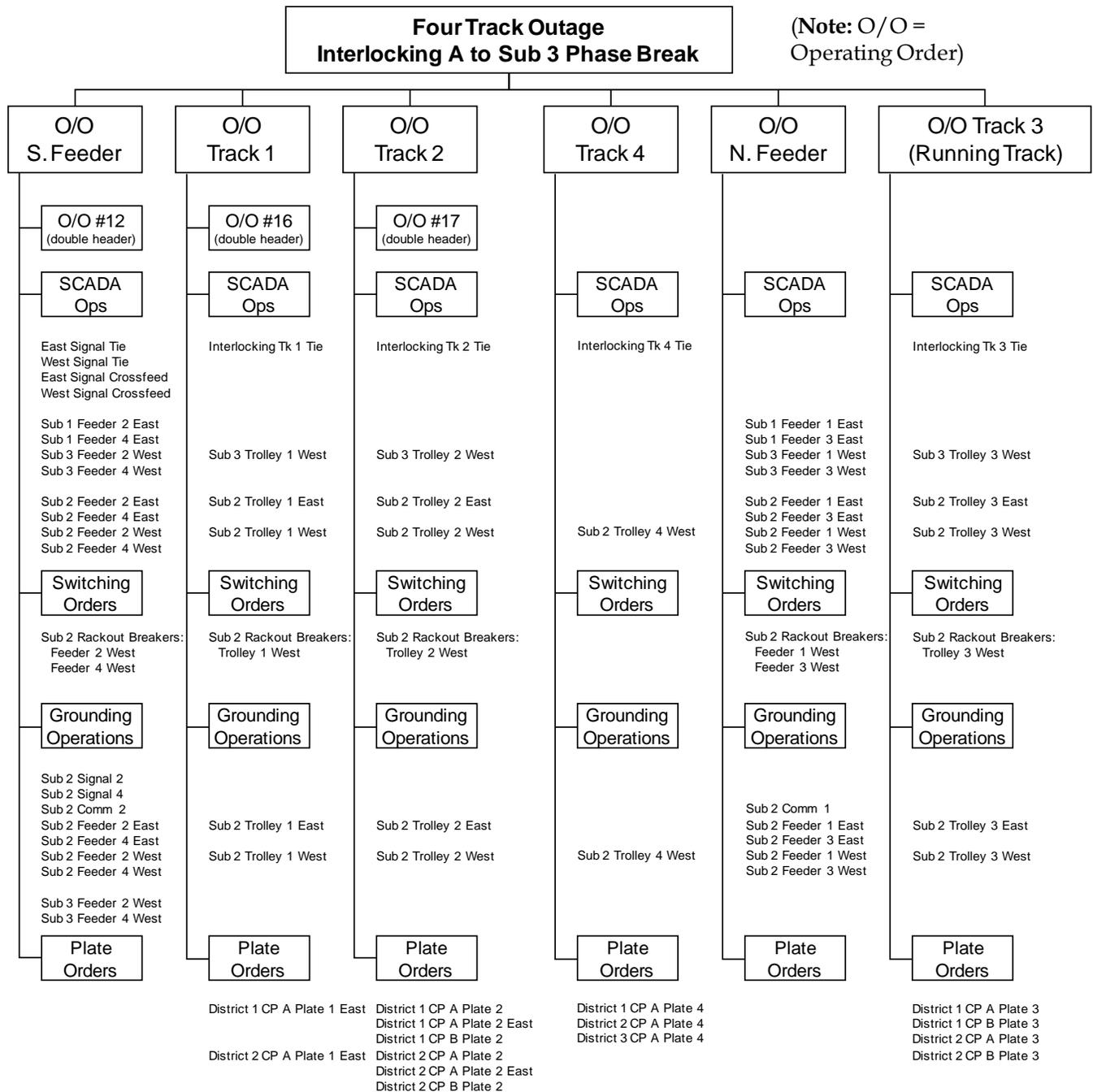

**Figure 4.** Box diagram of switching operations required for a typical four-track outage

Switching associated with each line group must be shown on a separate Power Director's Operating Order, which is in effect a checklist to ensure all steps of the outage procedure are properly followed. For this outage, there were six forms: the four tracks, and one each for the feeder line groups on each side of the right-of-way. On each Operating Order, the associated SCADA Operations & Tags, Switching Orders, Grounds, and Plate Orders are recorded. Figure 4 shows a simplified view of individual operations that must be commanded by the Power Director in a specific sequence. For this outage, 78 individual operations must be commanded (and another 78 operations executed to restore power after the outage is given up). Each operation can take between 30 seconds and 15 minutes, depending on whether the field Linemen are involved, and how far they have to walk or drive to reach the physical switch location.

## Results: Effective Work Window Available

How long does it really take to deliver a power outage, given all the necessary steps from the time when the Conductor Flag completes the track outage process (needed to apply local grounds to the catenary) until when power is isolated and a briefing given to Contractors regarding the allowable limits of work? This question should be answered statistically, but for this paper we use a single observation known to be representative of typical occurrences.

Figure 5 shows a typical weeknight when work is scheduled between 10 PM and 5 AM on one of the four tracks. Of seven hours overnight typically considered a "work window", due to the work location, service continued to run for 2 hrs 15 mins, followed by a half-hour spent removing tracks from service, 45 mins on remote switching operations, and 36 mins required to physically operate disconnects and establish grounds. It was 2:06 AM by the time Contractor's safety briefing was completed.

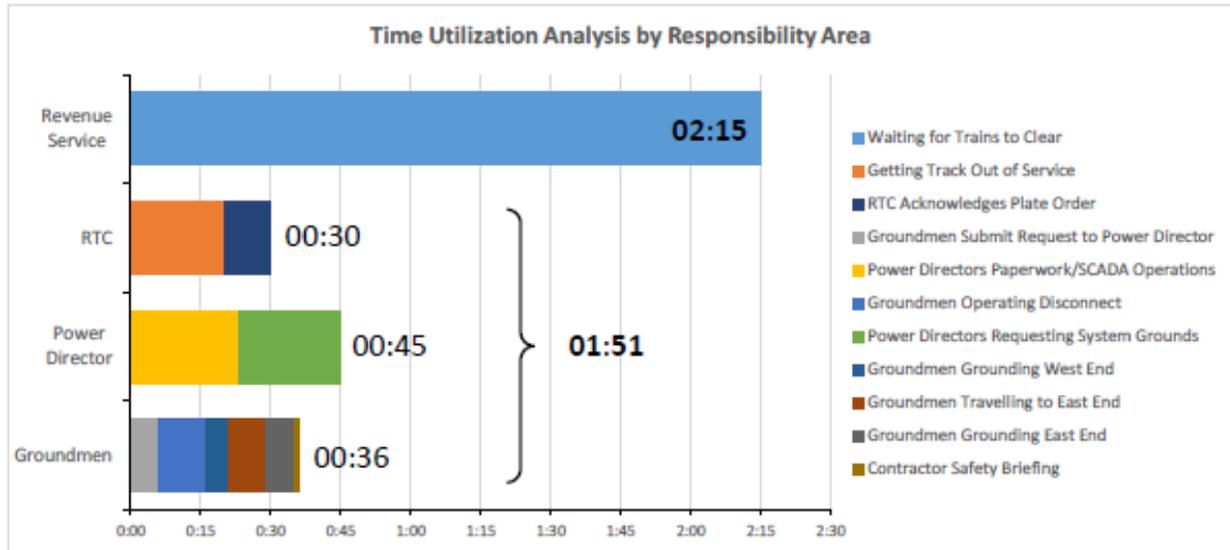

**Figure 5.** Electrical isolation process time utilization analysis

With the Dispatcher granting an extension of time, the Contractor worked until 4:45 AM before being told to clear up. Additional time was needed to reverse all circuit changes, allowing tracks to be opened to traffic at 5:30 AM. The total available work window that night (i.e. Contractor on-track time) was 2 hours and 39 minutes.

The Contractor planned on completing a trolley wire run that night. Despite a shorter work window than desired, they completed the planned segment of about 25 catenary lengths. On this particular night, work was complicated by an interlocking within the working limits. Had they been able to receive more on-track time, they might have been able to plan for, and complete, a longer section, and made more overall project progress. Figure 6 shows some of the necessary steps for this work.

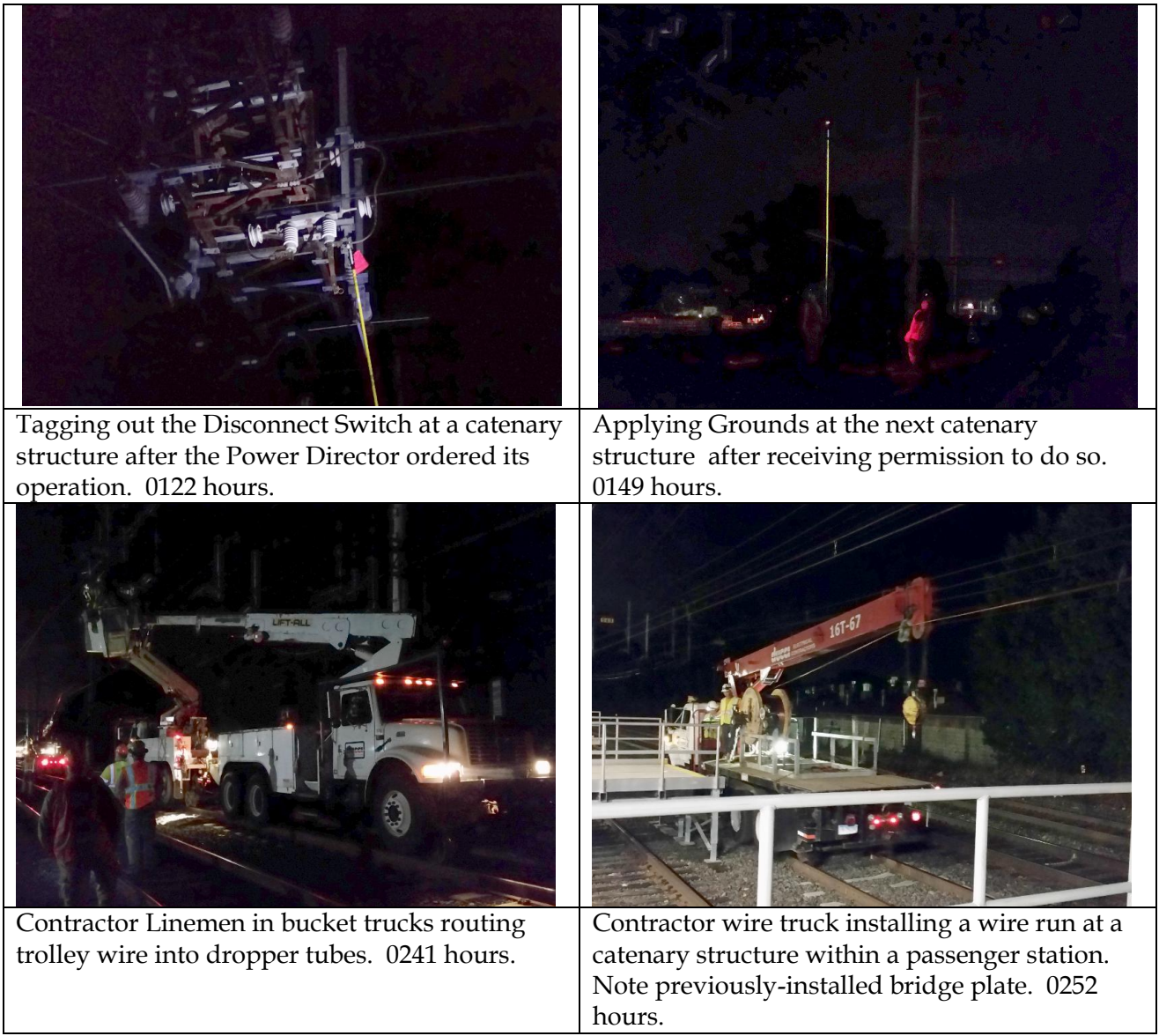

**Figure 6.** Night power isolation and catenary renewal work

Other electrified railways that chose to replace fixed-termination with auto-tensioning catenary have found the conversion process operationally disruptive. When an electrified line in the Midwest was undertaking this type of project, it closed 96-km (60 miles) of mainly single-track railroad on its suburban end to all traffic for multiple weekends [25]. This resulted in much longer work windows (and faster project completion). However, the case study segment is a busy four-track railway and a key segment of an interstate passenger and freight corridor, where such extensive planned closures are simply not feasible.

## Conclusions: Our Observations

Electric railways are fast, clean, safe, modern, noiseless... but very complicated to operate and maintain!  In high density, high-utilization corridors, where the benefits of electric traction are most evident, the complexity of ongoing maintenance and operations also threatens to impose significant costs.  What are some steps we can take to ensure electric traction infrastructure performs at its best?

- **Coordinate Electrical and Civil Design Elements with System Operations in Mind.**  Track layout is the domain of the civil engineer, where the priorities are track speed, capacity, and safety.  Power infrastructure is designed by the electrical engineer where the concerns are cost, load capability, and safety.  When possible, interlockings and traction substations should be co-located, or at least designed jointly, to simplify maintenance and emergency isolations.  This is not always possible, especially on 113-year-old electrical infrastructure that had been subject to numerous retrofits.  The operational challenges highlighted in this case study underscores the importance of maintenance considerations as design criteria.

- **Consider Cross-Training Between Disciplines.**  Those having a solid understanding of both electrical and track infrastructure, maintenance processes, and operating procedures are absolutely vital in both planning and execution of construction jobs.  Typically, Transportation, Track, and Power are different departments within a railroad, but cross-training is possible with union agreements or promotional paths that facilitate the acquisition and retention of such knowledge.  However, it can also have disadvantages in terms of higher pay rates and training costs, coupled with the risk of losing trained individuals.

- **Limiting Number of Concurrent Jobs Substantially Reduces Complexity.**  Overlapping power isolations are particularly disruptive when two outages overlap but do not begin and end at the same time, resulting in the need to re-configure circuits four times a weekend (for outage A, outages A+B, outage B, and to clear up).  Although having fewer concurrent jobs reduces complexity, it does not permit spreading the cost of electrical switching and alternate service over multiple maintenance tasks (called "piggybacking"). Trade-offs between cost-spreading and risk of complexity (based on realistically operable plans) is an important area for further research.

- **Formal Study of Maintenance Outage Capacity is Required.** We believe the maintenance outage capacity that could be sustained by the infrastructure should be assessed formally, separately for track and power infrastructure. There exists a maximum practical operational capacity to provide track and power outages whilst maintaining train service, which is something that may not have been fully appreciated by capital planners. This "outage capacity" should be assessed formally, similar to how track signalling capacity is assessed now (e.g. maximum throughputs in terms of trains per hour). At some service level, the outages that can realistically be provided will exceed those required for steady-state maintenance, and at that point we will essentially be "consuming" previously installed infrastructure to operate today's service. It can be tempting to assume that necessary work windows can be provided for "somehow," but in fact complexity of maintenance processes on legacy infrastructure requires active consideration. These findings form parts of constraints input to modern optimizing service planning models that maximizes asset utilization.

- **Automation in Several Areas Offers Potential Improvements.** Having described the operational process and current technology, we propose several areas of automation that could lead to long-term improvements

  1. **Higher Level of Automation in Power Switching Operations.** The philosophy of SCADA system design has focused on maximum flexibility at the expense of lesser degrees of automation. Train dispatching was like this during the mid-20th century, but Transportation officials have accepted that the reduced flexibility in plant operations have led to faster and safer train operations by bundling commonly used operations together (e.g., NX route setting). This concept could be further explored in power switching operations, especially in high density areas.

  2. **Further Research Could Focus on Work Window Optimization.** This operational challenge may be formulated as an optimizing program, where information about work requests, service needs, and resource availability is fed into the computer in real time. If based on rules about resource assignment and electrical infrastructure physical characteristics, the program could determine which work requests are feasible to support (including sharing of outages where possible). However, coordination would still be required between stakeholders to ensure that the latest information is entered, and the program must be capable of responding to changing circumstances, such as weather and personnel availability.

3. **Cognitive Artificial Intelligence (AI) Assistant Could Be Helpful.** While the typical philosophy for SCADA has been a focus on human operators performing the correct operating steps, a cognitive AI in a form of virtual assistant may be helpful. The nature of the SCADA system is relatively well defined in terms of the number of operations. With sufficient operational history, it should be possible to build a reinforcement learning system, which over time could suggest proper operations to the human operator and validate when improper system conditions are detected.

We hope this case study informs design engineers, transportation planners, public officials, and others contemplating the relative strengths and drawbacks of electric versus diesel traction for passenger railways. Electric traction has been seen as the gold standard for many years, and indeed it is an excellent technology for very high-density corridors or long-distance services requiring very high operating speeds. However, maintenance of such complex infrastructure imposes an ongoing cost not just in dollars but also operational inconvenience, and consumes managerial bandwidth. High-speed diesel locomotives or railcars can be an attractive option in areas of relatively lower demand density.

## Acknowledgments

The authors gratefully acknowledge the contributions of the following members of the Metro-North family in research assistance, education, mentoring, and making this work possible: J. Gillies, J. Balzano, J. Frank, W. Lockery, B. Kelly, G. Raguseo, A. Golino, D. Williard, M. Najam-ud-Din, C. Walters, L. Borsellino, M. Chen, Y. Yoshida, J. Pfeiffer, J. Pepitone, and J.E. Kesich. We'd also like to acknowledge the contribution of a number of external reviewers who wishes to remain anonymous. Responsibility for errors or omissions remains with the authors. All opinions expressed or implied are the authors' and do not necessarily reflect official policy or positions of the New York State Metropolitan Transportation Authority nor the Connecticut Department of Transportation.